\documentclass[preprint,12pt,onecolumn]{elsarticle}

\usepackage{epsfig}
\usepackage{fullpage}
\usepackage{amssymb,amsmath}
\usepackage{array}
\usepackage{graphicx,color,bm,amsmath,amssymb}
\usepackage{hyperref}
\usepackage{lineno}
\usepackage{graphicx}% Include figure files
\usepackage{dcolumn}% Align table columns on decimal point
\usepackage{bm}% bold math
\usepackage{color}
\usepackage{epstopdf}
\usepackage{gensymb}
\usepackage{caption}
\usepackage{subcaption}
\usepackage{upgreek}
\usepackage[export]{adjustbox}

\newcommand{\La}{\mathcal{L}}
\newcommand{\Da}{\mathcal{D}}

\newcommand\Pran{\ensuremath{\textit{Pr}}} % Prandtl number, cf TeX's \Pr product
\newcommand\Pe{\ensuremath{\textit{Pe}}}  % Peclet number
\newcommand\Kn{\ensuremath{\textit{Kn}}}  % Knudsen number
\newcommand\Ja{\ensuremath{\textit{Ja}}} 
\newcommand\Le{\ensuremath{\textit{Le}}}  
\newcommand\Bt{\ensuremath{\textit{B}_T}} 
\newcommand\Bm{\ensuremath{\textit{B}_M}}

\newcommand\tth{\ensuremath{\tilde{\theta}}}
\newcommand\tx{\ensuremath{\tilde{x}}}
\newcommand\ty{\ensuremath{\tilde{y}}}

\newcommand\tit{\ensuremath{\tilde{t}}}
\newcommand\tphi{\ensuremath{\tilde{\phi}}}

\newcommand\barho{\ensuremath{\bar{\rho}}}
\newcommand\bap{\ensuremath{\bar{p}}}
\newcommand\bak{\ensuremath{\bar{k}}}
\newcommand\badel{\ensuremath{\bar{\delta}}}
\newcommand\bacp{\ensuremath{\overline{c_p}}}
\newcommand\bavs{\ensuremath{\overline{v_s}}}
\newcommand\bat{\ensuremath{\overline{\delta T}}}
\newcommand\baa{\ensuremath{\bar{\alpha}}}

\begin{document}

\begin{frontmatter}

\title{{\bf Rate-limiting factors in thin-film evaporative heat transfer processes}}

\author[1]{H. Zhao\corref{cor1}}
\author[1]{R. Poole}
\author[2]{Z. Zhou}
\cortext[cor1]{\texttt{H.Zhao2@lboro.ac.uk}}
\address[1]{Wolfson School of Mechanical, Electrical and Manufacturing Engineering, 
Loughborough University, Loughborough, Leicestershire, LE11 3TU, UK}
\address[2]{State Key Laboratory of Multiphase Flow in Power Engineering, Xi'an Jiaotong University, Xi'an 710049, China}

\begin{abstract}
In this paper, we present a theoretical study aimed at investigating the rate-limiting factors in thin-film evaporative heat transfer processes, considering the finite-rate evaporation kinetics. The problems of evaporation of a flat thin-film in either pure vapours or vapour-inert-gas mixtures are analysed based on the non-dimensionalised macroscopic transport equations for continuum fluids, coupled with out-of-equilibrium kinetic boundary conditions. Both the full numerical solutions and asymptotic analytical solutions at slow evaporation limit are provided and applied to analyse thin water film evaporation. Existing solutions, assuming negligible heat transfer in the gas domain, or negligible temperature jump across the non-equilibrium kinetic layer, or more boldly a thermodynamically equilibrial interface (i.e. its temperature is at the saturation temperature), can be fully recovered from the more general solutions presented here. Our results show that while these assumptions hold in special cases, they can lead to significant errors in many conditions, especially when the film thickness $\delta$ is reduced to a few micrometers or thinner. We show that the conventional views that the rate-limiting factors in thin-film evaporative heat transfer is either the heat diffusion through the liquid film or the mass transfer in the gas domain only apply to thick film (i.e. $\delta \gg \lambda$ where $\lambda$ is the mean free path in the vapour phase). As $\delta$ decreases to a few micrometer or smaller (more precisely when the Knudsen number $\Kn$ increases beyond $O(1)$ in an pure vapour environment or when the kinetic Peclet number $\Pe$ is reduced below $O(1)$ in inert gases), the interfacial thermal resistance due to the evaporation kinetics can be on the same orders of magnitude as the thermal resistance in the liquid film. The analysis also allow us to compare the heat transfer processes during the evaporation of a thin-film in pure vapours to those in inert gases, providing deeper insight into the effectiveness of various strategies for exploring the evaporation process in practical thermal management.
\end{abstract}


\end{frontmatter}

\section{Introduction}
Cooling through evaporation has been explored widely in nature (e.g. perspiratory evaporation) and in industrial applications (e.g. boiling, heat pipes) due its high cooling efficiency \citep{Carey2020,Persad2016}. The theoretical maximum heat flux through evaporation at the liquid-vapour interface has also long been recognised to be orders of magnitude higher than what is achieved in practical evaporative cooling technology, e.g. critical heat flux in boiling \citep{Carey2020} or thin-film evaporation devices \citep{Vaartstra2020}. This discrepancy primarily stems from the significantly less efficient heat transfer process within the liquid film (i.e. between the hot surface and the liquid-vapour interface) or the gas domain. 

\begin{figure}[t]
\centering
\scalebox{1}{\includegraphics{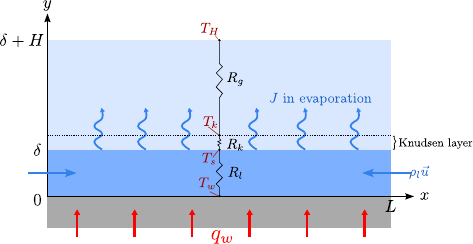}}
\caption{Schematic representation of a steady-state evaporation of a thin film and the thermal resistance network. The gray area represents the solid wall which is exposed to a uniform heat flux $q_w$. The dark blue area represents the thin film which is constantly replenished with mass flux $\rho_l \vec{u}$. Evaporation with mass flux $J$ occurs steadily on the interface. The light blue region represents the gas phase (can be either pure vapour or vapour-inert gas mixture). $R_g$ represents the thermal resistance in the gas phase. $R_k$ represents the thermal resistance during the evaporation and $R_l$ represents the thermal resistance in the liquid film. Thermal radiation is considered to be negligibly small.
\label{fig:RN}}
\end{figure}

Consider a simplified configuration where a flat thin film (comprising only species A with thickness $\delta$) evaporates steadily on a hot flat surface exposed to a uniform heat flux $q_w$. The heat transfer process can be visually represented by a resistance network along the thermal paths connecting the hot surface, the liquid domain, the interface region and the gas domain, as shown in Fig. \ref{fig:RN}, such that $q_w=(T_w-T_s)/R_l=(T_s-T_k)/R_k=(T_k-T_H)/R_g$, where $T$ is the temperature. Parameters evaluated at the wall, the interface, the outer boundary of the non-equilibrium Knudsen layer and the top boundary will be labelled with subscripts $w, s, k$ and $H$ respectively. In addition, any properties of the liquid domain will be labelled with a subscript $l$ and those for the gas domain with a subscript $g$.

$R_l$ represents the thermal resistance within the liquid domain, and is dominated by the thermal diffusion process in a sufficiently thin liquid film (see section \ref{sec:theory-vapour}) so $R_l=\delta/k_l$. $R_g$ represents the thermal resistance within the gas domain. It is commonly considered negligible in a pure vapour environment (i.e.\,consisting only of vapour phase of A) due to the fast pressure relaxation process at low Mach number \citep{Lu2019, Vaartstra2020}. However, in the case of a vapour-inert gas mixture (i.e.\,comprising vapour phase A and inert gases B), $R_g$ can be the rate-limiting resistance due to the slow mass diffusion process of the vapour in inert gases (i.e.\,diffusion limited)\citep{Lu2017, Vaartstra2020}. There is a general consensus in the literature that it is either $R_l$ or $R_g$ that constrains the overall heat transfer rate, preventing attainment of the theoretical maximum heat flux through evaporation \citep{Carey2020,Vaartstra2020}. To substantiate this claim, we need to quantify the magnitude of $R_k$. 

The evaporation process is fundamentally driven by the imbalance of the chemical potentials across the liquid-vapour interface. One of the most widely adopted model to predict the evaporation mass flux $J$ is the Hertz-Knudsen model \citep{Knudsen1950}, i.e. eq. \ref{eq:HK}, which predicts $J$ by counting the difference between the number of vapour molecules arriving at the interface and eventually condensing, and those of  molecules departing from the interface (i.e. evaporation).
\begin{equation}
J = \frac{1}{\sqrt{2\pi R_s^A}}\left(\sigma_e \frac{p_s^A}{\sqrt{T_s}} - \sigma_c \frac{p_k^A}{\sqrt{T_k}}\right).
\label{eq:HK}
\end{equation}
In this expression, $\sigma_e$ and $\sigma_c$ are empirical evaporation and condensation coefficients, commonly interpreted as the fraction of molecules which reach the interface and ultimately evaporate or condense. $p^A$ denotes the pressure of species A (or partial pressure in presence of inert gases). $R_s^A$ stands for the specific gas constant of species A. The assumptions of equilibrium Maxwellian velocity distribution in the gas phase and decoupled evaporation and condensation processes (i.e. $\sigma_e$ and $\sigma_c$ are constants) are taken implicitly. The HK relation is considered as semi-empirical, due to the absence of explicit expressions for $\sigma_e$ and $\sigma_c$. Analytical expressions for $\sigma_e$ and $\sigma_c$ based on the Statistical Rate Theory in the Thermal-Energy-Dominated limit (TED-SRT) were recently proposed by Persad and Ward \citep{Persad2016}. Their formulation demonstrating a strong coupling between the evaporation and condensation, i.e. both $\sigma_e$ and $\sigma_c$ are functions of $p_s^A,\,T_s,\,p_k^A$ and $T_k$. Several common assumptions are adopted routinely in the literature to simplify the analysis, including:
\begin{enumerate}
\item Sharp equilibrial interface assumption: the interface region is infinitely thin and $p_s^A$ can be taken as the saturation pressure at $T_s$ (needs correction based on the total pressure at the interface $p_s$ in presence of inert gases \citep{Zhao2023}). Analysis that counts for finite thickness and intermolecular forces indicate that the interface region in the liquid side is indeed close to equilibrium \citep{Benilov2023}.
\item Negligible temperature jump across the Knudsen layer (i.e. $T_s \approx T_k$) \citep{Carey2020,Lu2019,Sultan2005a}.
\end{enumerate}

The second assumption can lead to the conclusion that $R_k$ is infinitely small and can be ignored. Please note that, although $R_k \rightarrow 0$, the achievable heat flux can still strongly depend on the evaporation kinetics because of the temperature-dependent finite evaporation rate, i.e.\,the value of $T_s$ is kinetics dependent. The only cases when the heat transfer rate is independent of the evaporation kinetics is when the interface is at thermodynamic equilibrium (i.e. $T_s = T_k = T_{sat}(P_k^A)$). The local thermodynamic equilibrium assumption essentially assumes that the evaporation rate is infinitely large to bypass the kinetic theory and is a widely adopted in problems dealing with evaporation in inert gases \citep{Lu2017, Sirignano2010}.  As a result, in cases when the temperature jump is negligible, the way to reduce the overall thermal resistance in a pure vapour environment is to keep reducing the thickness of the liquid film. It is well known that as the film thickness is reduced to be comparable to the distance where the attractive intermolecular force between the solid surface and the interface becomes prominent (e.g. $< 100$ nm \citep{Craster2009}), the evaporation will be further suppressed by the Casimir force and the above formulation needs to be corrected by counting for the disjoining pressure \citep{Chen2022, Craster2009}. In addition, the intermolecular forces can also contribute to destabilize and break the nanometers-thick-film \citep{Craster2009,Zhao2020}.

In recent years, the assumption of negligible temperature difference across the Knudsen layer has been proved inaccurate through both experiments \citep{Persad2016} and analysis (e.g. solving the Boltzmann equation \citep{Struchtrup2017, Chen2022} or doing molecular dynamics simulations \citep{Hoyst2013a}). In a pure vapour environment,  a more generalised kinetic model capable of predicting both $J$ and the temperature jump has been derived by Struchtrup et al.\citep{Struchtrup2017}. This model is based on the perturbation analysis of the Boltzmann equation, accurate up the order of $\Kn^3$, where $\Kn=\lambda^A/\delta$ is the Knudsen number with $\lambda^A$ being the mean free path of gaseous species A. The model also assumes that the Knudsen layer is not too far from thermodynamic equilibrium, allowing the velocity profile in the gas domain to approximate a Maxwellian distribution. A similar model has also been derived recently by Chen \citep{Chen2022} which introduced a third correlation to predict the density jump across the Knudsen layer. Through these model and experiments, it has been shown that the magnitude of the temperature jump increases with the increase in $J$ \citep{Persad2016, Struchtrup2017, Chen2022, Chen2024}. This implies that the argument of negligible $R_k$ becomes more questionable at increasingly small $\delta$ where $J$ is significant. This raises the question that whether or when $R_k$ can become the rate-limiting factors during thin-film evaporation process, either in a pure vapour environment or in inert gases. Addressing this question forms the primary focus of this work, which involves a detailed analysis of the heat-mass transfer process depicted in fig.\ref{fig:RN}.

The theoretical framework used in this work will be outlined in section \ref{sec:theory}. The framework will couple the established macroscopic transport equations for continuum fluids (within both the liquid and gas domains) with kinetic boundary conditions. A similar framework has been recently employed by Chen \citep{Chen2024} to quantify the temperature jump in thin-film evaporation in a pure vapour environment, and by Zhao et al. \citep{Zhao2023} to study the evaporation of spherical droplet in inert gases. In this work, we will leverage to analyse the entire heat and mass transport processes depicted in fig.\ref{fig:RN}. Moreover, the analysis will adopt a dimensionless framework to reduce the number of variables, delineate conditions where different commonly-adopted assumptions hold, and identify the corresponding rate-limiting factors. Linearised asymptotic solutions at the slow evaporation limit will also be provided to get a more intuitive understanding of the evaporation process, which can also serve as significantly simplified explicit approximated solutions. Both the numerical and analytical asymptotic solutions of the problem will then be applied to examine the evaporation of a thin water film. The results will be presented and discussed in details on section \ref{sec:Results} before drawing conclusions in section \ref{sec:Conclusion}.

\section{Theory of steady evaporation of a flat thin-film}\label{sec:theory}

\subsection{Evaporation in pure vapour} \label{sec:theory-vapour}
Consider a simplified two-dimensional flow which is invariant along the $z$-axis, the temperature distribution within the thin liquid domain, characterised by constant thermo-physical properties, can be analysed based on the following conservation of energy:
\begin{equation}
\frac{\partial \theta}{\partial t} + u \frac{\partial \theta}{\partial x} + v \frac{\partial \theta}{\partial y} = \alpha_l \left(\frac{\partial^2 \theta}{\partial x^2}+\frac{\partial^2 \theta}{\partial y^2}\right),
\label{eq:El}
\end{equation}
where $u$ and $v$ are the velocity along the $x$ and $y$ direction respectively, $\alpha$ is the thermal diffusivity, $\theta = T - T_H$, and $T_H$ is the temperature at $y=\delta+H$.\\

Rescale eq. \ref{eq:El} using the following characteristic parameters: $\theta^\star = T_s-T_H, x^\star = L, y^\star = \delta, u^\star = JL/(\rho_l \delta), v^\star = J/\rho_l, t^\star = \delta/v^\star = \rho_l \delta/J$, we can get
\begin{equation}
\baa \barho \Lambda\left(\frac{\partial \tth}{\partial \tit} + \frac{\partial \tth}{\partial \tx} + \frac{\partial \tth}{\partial \ty}\right) = \left(\frac{\delta}{L}\right)^2 \frac{\partial^2 \tth}{\partial \tx^2}+\frac{\partial^2 \tth}{\partial \ty^2},
\label{eq:El_D}
\end{equation}
where $\baa = \alpha_g/\alpha_l, \barho = \rho_g/\rho_l$, and all parameters with a tilde sign are dimensionless variables with magnitude of O(1). $\Lambda = J\delta/(\rho_g \alpha_g)$ can be defined as the evaporation number \citep{Sultan2005a,Kanatani2013} and is one of the key dimensionless numbers in this evaporation problem. From eq.\ref{eq:El_D}, it is clear that at the limit of thin film (i.e. $\delta/L \ll 1$) and not so fast evaporation (i.e. $\baa \barho \Lambda \ll 1$), a non-convective diffusion process dominates which gives rise to a linear temperature profile within the liquid film such that $q_w = k_l (T_w - T_s)/\delta$.

In a steady-state incompressible gas domain with a small Rayleigh number ($Ra = g\beta_g\theta^\star H^3/(\alpha_g \nu_g)$, where $g$ is the gravitational acceleration, $\beta_g$ is the thermal expansion coefficient, and $\nu_g$ is the kinematic viscosity), the continuity equation requires $u \approx 0, v = J/\rho_g$ and the heat transfer process can be modelled by the following one-dimensional energy equation:
\begin{equation}
v \frac{d\theta}{dy} = \alpha_g \left(\frac{d^2 \theta}{dy^2}\right).
\label{eq:Ev}
\end{equation}
Solving eq.\ref{eq:Ev} requires two boundary conditions: one at $y = \delta + H$ where $\theta = 0$ and the other at $y = \delta$ where the conservation of energy across the interface requires:
\begin{equation}
k_l \frac{\theta_w - \theta_s}{\delta} = q_v + J\La, \quad q_v = - k_g \left[\frac{d\theta}{dy}\right]_k,
\label{eq:BC2}
\end{equation}
where $\La$ is the latent heat of evaporation and $q_v$ is the heat flux added to the vapour. $q_v$ is commonly assumed to be negligible in the literature, and its contribution has been modelled in Chen's latest work \citep{Chen2024}. This boundary condition requires other enclosure equations to model $J$ which depends on the evaporation kinetics. Here, we adopted the model derived by Struchtrup et al. \citep{Struchtrup2017} in forms of the extension of the Onsager-Casimir reciprocity relations:
\begin{equation}
\left[ \begin{array}{c} 
\frac{p_s^{eq}-p_k}{\sqrt{2 \pi R_s^A T_s}} \\ \frac{p_s^{eq}}{\sqrt{2 \pi R_s^A T_s}} \bat
\end{array} \right] 
= \left[ \begin{array}{cc}
r_{11} & r_{12} \\ r_{21} & r_{22}
\end{array}\right]
\left[ \begin{array}{c} 
J \\ \frac{q_v}{R_s^A T_s}
\end{array} \right],
\label{eq:R13}
\end{equation}
where $\bat = (T_s-T_k)/T_s$, coefficients $r_{11}=1/\pi+9/32$, $r_{12}=r_{21}=1/16+1/(5\pi)$, $r_{22}=1/8+13/(25\pi)$, $p_s^{eq}$ is the equilibrium vapour pressure of species A. The above expression is an extension of the HK model taking into account the Stefan flow due to net evaporation, with the additional assumption $\sigma_e=\sigma_c=1$. It also predicts the temperature jump across the Knudsen layer, which is commonly ignored in the literature. As shown by Persad and Ward \citep{Persad2016}, assuming either $T_k = T_s$ or $\sigma_e = \sigma_c$ does not introduce major discrepancies compared to experimental results ($\leq 5\%$), but making both assumptions simultaneously forces the Knudsen layer into equilibrium which could lead to significant errors. Chen \citep{Chen2024} also clearly shows that the temperature jump is a generic feature during the phase change process and is due to the mismatch of the enthalpy carried by vapour molecules at the interface and the bulk gas region.

After redefining $v^\star=J/\rho_g$ in the gas domain, eqs. \ref{eq:Ev} - \ref{eq:BC2} can be rescaled to
\begin{equation}
\Lambda \frac{d\tth}{d\ty} = \left(\frac{d^2 \tth}{d\ty^2}\right).
\label{eq:Ev_D}
\end{equation}
\begin{align}
\mathrm{at} \quad & \ty = \frac{H}{\delta}+1 =\frac{1}{\badel}+1: \quad \tth = 0,\\
\mathrm{at} \quad & \ty = 1: \Ja (\tau - 1) - \Bt = -\Bt\bak\left[\frac{d\tth}{d\ty}\right]_k+\Lambda\bak, \\
 & \qquad \quad \& \quad \tth_k = 1 + \bat \left(\frac{1}{\kappa} + 1\right),
\label{eq:BC2_D}
\end{align}
where
\begin{equation*}
\Ja = \frac{c_p^A T_H}{\La},\quad \Bt = \frac{c_p^A \theta^\star}{\La},\quad \tau = \frac{T_w}{T_H},\quad \bak = \frac{k_g}{k_l},\quad \kappa=\frac{\Bt}{\Ja}.
\end{equation*}
$\Bt$ is the Spalding number of heat transfer. Similarly, eq.\ref{eq:R13} can also be rescaled to
\begin{gather}
\frac{2\Lambda\Kn\sqrt{1+\kappa}}{\Pran}=\hat{r}_{11}[\bap \exp(\chi)-1]+\hat{r}_{12} \bat, \label{eq:R13_1D}\\
\frac{\bat \Pran \sqrt{1+\kappa}}{\Kn} = 2 r_{11} \Lambda(1+\kappa) + 2 r_{12}\frac{\gamma}{\gamma-1}\left(\frac{\tau-\kappa-1}{\bak} - \frac{\Lambda}{\Ja}\right),
\label{eq:R13_2D}
\end{gather}
where $\hat{r}_{11}=r_{22}/(r_{11}r_{22}-r_{12}^2)$, $\hat{r}_{22}=-r_{12}/(r_{11}r_{22}-r_{12}^2)$, $\gamma$ is the heat capacity ratio, and 
\begin{equation*}
\Kn = \frac{\lambda}{\delta},\quad \lambda = \sqrt{\frac{\pi}{2 R_s^A T_H}}\nu_g,\quad \Pran = \frac{\nu_g}{\alpha_g},\quad \bap = \frac{p_{sat}(T_H)}{p_H},\quad \chi = \frac{\gamma}{\gamma-1}\left(\frac{1}{\Ja} - \frac{1}{\Bt+\Ja}\right).
\end{equation*}
Eq.\ref{eq:R13_1D} has adopted the Clausius-Clapyron equation to calculate $p_s^{eq}$ with respect to the reference equilibrium condition \{$T_H, p_{sat}(T_H)$\}. It is clear from eqs.\ref{eq:R13_1D} and \ref{eq:R13_2D} that the classical assumptions of negligible temperature jump and/or infinitely fast evaporation kinetics are only valid at the small $\Kn$ limit (i.e. $\Kn \rightarrow 0$ or equivalently $\delta \rightarrow \infty$). As the film gets thinner, both the temperature jump and finite-rate evaporation kinetics become important and needs to be accounted for, as shown by Chen \citep{Chen2024}.

Eqs.\ref{eq:Ev_D} - \ref{eq:R13_2D} can then be solved together to get the following more general conservation of energy across the interface:
\begin{equation}
\Ja (\tau - 1) - \Bt = \frac{\bak \Lambda}{\exp(\Lambda/\badel)-1}(\Bt+\Ja\,\bat)+\Lambda\bak.
\label{eq:EI_D}
\end{equation}
The full solutions (i.e. \{$\Lambda, \Bt \,\&\, \bat$\}, or equivalently \{$J, T_s \,\&\, T_k$\} in the physical space) can then be obtained by solving eqs.\ref{eq:R13_1D}, \ref{eq:R13_2D}, and \ref{eq:EI_D} numerically together.

The special cases of negligible $q_v$ can be solved by ignoring the first term on the right-hand-side of eq.\ref{eq:EI_D} and this will be a good approximation when $\{\Bt\,\&\,\bat\} \rightarrow 0$. In this special case, the transport process in the gas domain becomes irrelevant which leads to the so-called ``one-sided" model which is also widely adopted in the literature (see review in \citep{Craster2009}). In cases where the temperature jump is not too significant, the following asymptotic solution can be obtained by linearizing eqs.\ref{eq:R13_1D}, \ref{eq:R13_2D}, and \ref{eq:EI_D} about the equilibrium condition, i.e. $\{\Bt \, \& \,\bat\}=0$:
\begin{equation}
\left[ \begin{array}{c} 
\Bt \\ \bat
\end{array} \right] 
= \frac{1}{s_{11}^{HK}s_{12}^{IT} - s_{12}^{HK}s_{11}^{IT}}\left[ \begin{array}{c}
s_{12}^{HK}s_{10}^{IT} - s_{12}^{IT}s_{10}^{HK} \\ s_{11}^{IT}s_{10}^{HK} - s_{11}^{HK}s_{10}^{IT}
\end{array}\right], \quad \Lambda = \frac{1}{\bak}[\Ja(\tau-1)-\Bt],
\label{eq:OS}
\end{equation}
where
\begin{gather*}
s_{10}^{HK} = \frac{2\Kn\Ja(\tau-1)}{\bak\Pran}-\hat{r}_{11}(\bap-1);\quad s_{11}^{HK} = \frac{\Kn(\tau-3)}{\Pran\bak}-\frac{\gamma}{\gamma-1}\frac{\hat{r}_{11}\bap}{\Ja^2};\quad s_{12}^{HK}=-\hat{r}_{12};\\
s_{10}^{IT} = \frac{2r_{11}\Ja(\tau-1)}{\bak};\quad s_{11}^{IT} = \frac{2r_{11}(\tau-2)}{\bak};\quad s_{12}^{IT}=-\frac{\Pran}{\Kn}.
\end{gather*}
The full numerical solutions assuming negligible temperature jump can be obtained by setting $\bat=0$ and the corresponding asymptotic solution is: $\Bt = -s_{10}^{HK}/s_{11}^{HK}$. The classical solutions assuming thermodynamic equilibrium at the interface can also be recovered by taking $\Kn \rightarrow 0$. The general full numerical solutions together with existing solutions taking different assumptions will be compared and discussed in section \ref{sec:Results}.

Utilizing this dimensionless framework, we can also derive that
\begin{equation}
 \frac{R_l}{R_{ig}}=\frac{\Ja(\tau-1)}{\Bt} - 1, 
\end{equation}
where $R_{ig}$ is the combined thermal resistance across the interface and gas domain and is used considering the interface kinetics will be strongly coupled to the transport process in the gas domain. It is clear that the negligible thermal resistance across the interface and gas domain corresponds to the limit $\Bt \rightarrow 0$.
\subsection{Evaporation in inert gases} \label{sec:theory-inert}
When there is insoluble inert gas B in the gas domain, the diffusion process between species A and B becomes important. The conservation of mass still requires $v=J/\rho_g$. As a result, the macroscopic transport equations in the gas domain consist of:
\begin{gather}
v \frac{d\phi}{dy} = \Da_g \left(\frac{d^2 \phi}{dy^2}\right), \label{eq:Dg}\\
\frac{c_p}{c_{p,H}} v \frac{d\theta}{dy} = \alpha_g \left(\frac{d^2 \theta}{dy^2}\right) + \Da\frac{\Delta c_p}{c_{p,H}}\frac{d\omega}{dy}\frac{d\theta}{dy}, \label{eq:Eg}
\end{gather}
where $\phi = \omega - \omega_{H}$, $\omega$ is the mass fraction of species A, $\Da$ is the mass diffusivity between species A and B, and $\Delta c_p = c_p^A - c_p^B$. Please note that eq.\ref{eq:Eg} ignores the viscous dissipation (i.e. when the Eckert number $Ec=\Da_g^2/(\delta^2 c_{p,H}\theta^{\star}) \ll 1$), as in eq.\ref{eq:Ev} and its last term counts for the energy exchange due to mass diffusion. The required boundary conditions to solve these transport equations are:
\begin{align}
\mathrm{at} \quad & y = H + \delta: \,\theta = 0,\quad \phi = 0; \label{eq:BC1_g}\\
\mathrm{at} \quad & y = \delta: k_l \frac{\theta_w - \theta_s}{\delta} = - k_g \left[\frac{d\theta}{dy}\right]_k + J\La, \label{eq:BC4_g}\\
 & \qquad \quad \& \quad J = \frac{\rho_g\Da[d\phi/dy]_k}{\phi_k}=\frac{1}{\sqrt{2\pi R_s^A T_s}}\left\{\hat{r}_{11}[\bap\exp(\chi+\bavs) - p_k^A]+\hat{r}_{12}p_k^A\bat\right\}, \label{eq:BC2_g}\\
 & \qquad \quad \& \quad \frac{T_k-T_s}{T_H - T_s}=\frac{1}{1+\delta/(c\lambda^B)}, \label{eq:BC3_g}
\end{align}
where $\lambda^B$ is the mean free path of inert gases and $\bavs$ corrects for change of chemical potential due to presence of inert gases and its expression assuming ideal gas is \citep{Zhao2023}:
\begin{equation*}
\bavs = \frac{1}{\rho_l}\left[\frac{p_s}{R_s^A T_s}-\frac{p_{sat}(T_s)}{R_s^A T_s}\right].
\end{equation*}
The mixed-kinetic-diffusion boundary condition shown in Eq.\ref{eq:BC2_g} was proposed and has been successfully applied to model the evaporation of a spherical droplet in inert gases \citep{Zhao2023}. In the same work, the temperature jump condition without counting for the inert gases, i.e. eq.\ref{eq:R13_2D}, was shown to be not valid while the empirical temperature jump condition (i.e. eq. \ref{eq:BC3_g} with $c=2.35$) which is obtained by fitting to results from molecular dynamic simulations \citep{Hoyst2013a} provided a much better prediction of the evaporation process. While eq. \ref{eq:BC3_g} demonstrates the dominant role of inert gases in the temperature jump prediction, it is empirical and does not count for energy exchange which contributes to the difference in enthalpy and consequentially the temperature jump. A more rigorous kinetic model based on the energy exchange while counting for the interactions between the vapour phase of A and inert gases is still absent in existing literature and needs to be developed in the future.

Introducing an additional characteristic parameters: $\phi^\star = \omega_k - \omega_H$, the full problem can be rescaled to:
\begin{gather}
\lambda \Le \frac{d\tphi}{d\ty} = \left(\frac{d^2 \tphi}{d\ty^2}\right), \label{eq:Dg_D}\\
(1+\Pi\tphi)\Lambda \frac{d\tth}{d\ty} = \frac{d^2 \tth}{d\ty^2} + \frac{\Pi}{\Le}\frac{d\tphi}{d\ty}\frac{d\tth}{d\ty}, \label{eq:Eg_D}
\end{gather}
where 
\begin{equation*}
\Le = \frac{\alpha_g}{Da},\quad \Pi = \frac{\Bm(\bacp-1)}{\Bm+1}, \quad \bacp = \frac{c_{p,H}^A}{c_{p,H}},\quad \Bm=\frac{\phi^\star}{1-\omega_k}.
\end{equation*}
$\Le$ is the Lewis number, $c_{p,H}$ is the heat capacity of the gas mixture at the top boundary, and $\Bm$ is the Spalding number of mass transfer. These two transport equations are subject to the following rescaled boundary conditions:
\begin{align}
\mathrm{at} \quad & \ty = \frac{1}{\badel} + 1: \,\tth = 0,\quad \tphi = 0; \label{eq:BC1_g_D}\\
\mathrm{at} \quad & y = 1: \Ja (\tau - 1) - \Bt = -\Bt\bak\left[\frac{d\tth}{d\ty}\right]_k+\Lambda\bak\bacp,\quad \tth_k = 1 + \bat \left(\frac{1}{\kappa} + 1\right)\label{eq:BC4_g_D}\\
 & \qquad \quad \& \quad \frac{\Lambda\Le\sqrt{1+\kappa}}{\Pe}=-\frac{\Bm\sqrt{1+\kappa}}{\Pe}\left[\frac{d\tphi}{d\ty}\right]_k \notag \\
 & \qquad \qquad \qquad  = \hat{r}_{11} \bap \exp(\chi+\bavs)\left[1+\frac{(\varepsilon-1)\Phi_H}{\Bm+1}\right]-\left(1-\frac{\Phi_H}{\Bm+1}\right)(\hat{r}_{11}-\hat{r}_{12} \bat), \label{eq:BC2_g_D}\\
 & \qquad \quad \& \quad \bat=-\frac{\kappa}{(1+\kappa)[1+1/(\Kn^B c)]}, \quad \Kn^B = \frac{\lambda^B}{\delta} \label{eq:BC3_g_D}
\end{align}
where 
\begin{equation*}
\Pe = \frac{u_k \delta}{\Da},\quad u_k = \sqrt{\frac{R_s^A T_H}{2\pi}}, \quad \bavs = \frac{\barho}{\kappa+1}(\varepsilon\Phi_H+1-\Phi_H)[1-\bap\exp(\chi)],\quad \varepsilon=M^A/M^B,
\end{equation*}
and $\Phi = 1 - \omega, \bap = \rho_g/\rho_l$. $\Pe$ is the kinetic Peclet number which plays a similar dominant role as $\Kn$ in pure vapour, and $M$ is the molecular mass. As $\Pe \rightarrow \infty$, the local thermodynamic equilibrium assumption holds well. As $\Kn^B \rightarrow 0$, the temperature jump becomes negligible. Both scenarios occurs as $\delta$ increases. 

The solution to eq.\ref{eq:Dg_D} together with eq.\ref{eq:BC2_g} leads to an interesting correlation: $\Bm=\exp(\Lambda\Le/\badel)-1$, and the following important equation:
\begin{equation}
\Lambda\Le = \ln(\Bm+1)\badel.
\label{eq:Dg_s}
\end{equation}
Combing eq.\ref{eq:Dg_D} and \ref{eq:Eg_D} also gives rise to a new simplified energy equation:
\begin{equation}
\Lambda \bacp \frac{d\tth}{d\ty} = \frac{d^2 \tth}{d\ty^2}. \label{eq:Eg_D2}
\end{equation}
The solution to eq.\ref{eq:Eg_D2} together with eq.\ref{eq:BC4_g_D} give raise to the second important equation:
\begin{equation}
\Ja (\tau - 1) - \Bt = \frac{\bak \Lambda \bacp}{(\Bm+1)^{\bacp/\Le}-1}[\Bt+\bat(\Ja+\Bt)]+\Lambda\bak\bacp.
\label{eq:Eg_s}
\end{equation}
The full solution, i.e. \{$\Lambda, \Bt, \Bm, \bat$\} or equivalently \{$J, T_s, \omega_k, T_k$\}, can then be obtained by solving eqs.\ref{eq:BC2_g_D}, \ref{eq:BC3_g_D}, \ref{eq:Dg_s} and \ref{eq:Eg_s} simultaneously. The asymptotic solution at the slow evaporation limit can then be obtained by linearising these four equations about the condition where $\{\Bt, \Bm, \bat\}\rightarrow 0$, and we can get the following solution:
\begin{equation}
\begin{split}
\Bm & = \frac{(s_{22}^{HK}-s_{23}^{HK}s_{2}^{TJ})s_{24}^{T} + (s_{2}^{TJ}s_{23}^{T}-s_{22}^{T})s_{24}^{HK}}{(s_{22}^{HK}-s_{23}^{HK} s_{2}^{TJ})s_{21}^{T} - (s_{22}^{T}-s_{2}^{TJ}s_{23}^{T})(s_{21}^{HK}+\badel/\Pe)}, \\ 
\Bt & = \frac{s_{21}^{T}s_{24}^{HK} - (s_{21}^{HK}+\badel/\Pe)s_{24}^{T}}{(s_{22}^{HK}-s_{23}^{HK} s_{2}^{TJ})s_{21}^{T} - (s_{22}^{T}-s_{2}^{TJ}s_{23}^{T})(s_{21}^{HK}+\badel/\Pe)},\\
\bat & = - s_{2}^{TJ}\Bt,\quad \Lambda = \frac{\badel}{\Le}\Bm,
\end{split}
\label{eq:AS_f}
\end{equation}
where
\begin{gather*}
s_{21}^{HK} = \Phi_H \hat{r}_{11}[\bap\exp(d)(\varepsilon-1)+1], \quad s_{22}^{HK} = -\exp(d)\hat{r}_{11}\bap\left(\frac{\gamma}{(\gamma-1)\Ja^2}-f\right)[1+(\varepsilon-1)\Phi_H];\\
s_{23}^{HK} = \hat{r}_{12}(\Phi_H -1); \quad s_{24}^{HK}=\hat{r}_{11}\{\bap \exp(d)[1+(\varepsilon-1)\Phi_H]-1+\Phi_H\};\\
d = \barho[1+(\varepsilon-1)\Phi_H](1-\bap); \quad f = \barho[1+(\varepsilon-1)\Phi_H]\left[\frac{\bap \gamma}{(\gamma-1)\Ja^2}+\frac{1-\bap}{\Ja}\right];\\
s_{21}^{T} = \frac{\bak \badel \bacp}{Le};\quad s_{22}^{T}=\bak \badel; \quad s_{23}^{T} = \bak \badel \Ja; \quad s_{24}^{T} = \Ja(\tau-1); \quad s_{2}^{TJ} = \frac{1}{\Ja[1+1/(\Kn^B c)]}.
\end{gather*}
In the special cases when $q_v$ is negligibly small, the conservation of energy across the interface, i.e. eq.\ref{eq:Eg_s}, can be simplified as:
\begin{equation}
\Ja (\tau - 1) - \Bt = \Lambda\bak\bacp.
\label{eq:Eg_s_1}
\end{equation}
Eq.\ref{eq:Eg_s_1} together with eqs.\ref{eq:BC2_g_D}, \ref{eq:BC3_g_D} and \ref{eq:Dg_s} can be used to solve the problem. The linearised solutions are eq.\ref{eq:AS_f} but with $s_{22}^{T}=1$ and $s_{23}^{T}=0$. If the temperature jump is also negligibly small, $s_{2}^{TJ}=0$. The further assumption of thermodynamic equilibrium at the interface can be counted for by taking the limit $\Pe \rightarrow \infty$.

\section{Results and discussion} \label{sec:Results}
To investigate the characteristics and rate-limiting factors during an evaporation of a thin liquid film. The theory detailed in section \ref{sec:theory} is employed to model the evaporation of a thin water film in two different gas environments: 
\begin{enumerate}
\item pure water vapours where the top boundary is at thermodynamic equilibrium: $p_H = 1$ bar, $T_H=T_{sat}(p_H)=372.76$ K;
\item mixture of water vapour and gaseous nitrogen at different $\Phi_H, T_H$ and $p_H$.
\end{enumerate} 
All thermo-physical properties of water and nitrogen are evaluated at $T_H$ and $p_H$ and are obtained from the NIST database \citep{KroenleinK2012}. The thermal conductivity of the gaseous mixture is calculated based on the molar fraction weighted sum and the heat capacity is evaluated based on the mass fraction weighted sum. The wall temperature $T_w$ is fixed at 422.76 K (i.e. 50 K above $T_{sat}$ and 1 bar). 

Results showing the heat and mass transfer processes for a thin-film evaporating in its own vapour will firstly be presented and discussed in this section. The pure vapour case can also be considered as the limit of negligible inert gases in a more general vapour-inert-gas mixture gas domain (i.e. $\Phi_H \rightarrow 0$) and the solutions for the more general problems of evaporation in inert gases will subsequently be provided.
\subsection{Evaporation of water film in water vapour} \label{sec:Results-vapour}
\begin{figure}[t]
\centering
\begin{subfigure}{.32\textwidth}
\centering
\includegraphics[width = \textwidth]{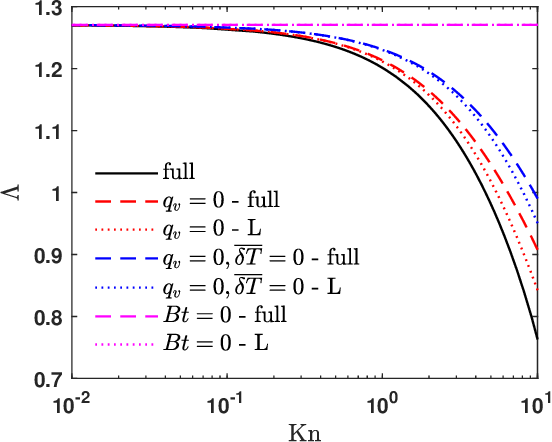}
\caption{ }
\label{fig2a}
\end{subfigure}
\hfill
\begin{subfigure}{.32\textwidth}
\centering
\includegraphics[width = \textwidth]{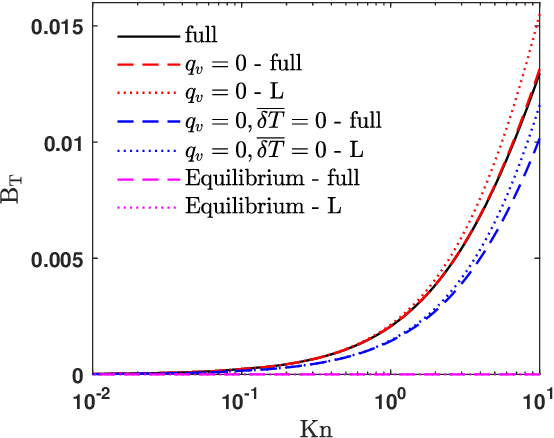}
\caption{ }
\label{fig2b}
\end{subfigure}
\hfill
\begin{subfigure}{.32\textwidth}
\centering
\includegraphics[width = 1.05\textwidth]{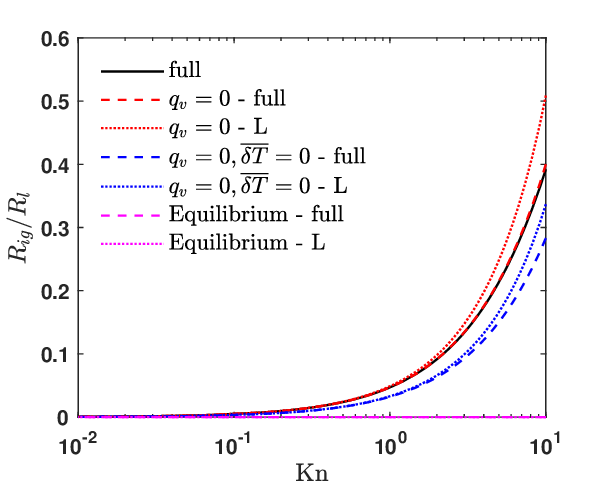}
\caption{ }
\label{fig2c}
\end{subfigure}
\caption{Impact of $\Kn$ on (a) $\Lambda$, (b) $\Bt$ and (c) $R_ig/R_l$, under different assumptions, including negligible heat transfer in the vapour domain (i.e.\,$q_v = 0$), negligible temperature jump across the Knudsen layer (i.e. $\bat = 0$) or interface is at thermodynamic equilibrium (i.e.\,$T_s=T_{sat}$). In all cases, $\badel=1,\,T_H = T_{sat}(p_H)=372.76 \mathrm{K}, p_H=1$ bar. `full' - full solutions; `L' - linearised solutions. }
\label{fig2}
\end{figure}

The impact of various assumptions on the predicted $\Lambda$ and $\Bt$ at different $\Kn$ is shown on fig.\ref{fig2a} and \ref{fig2b}. As expected, the classical assumption of local thermodynamic equilibrium (i.e.\,two pink lines) implies infinitely fast evaporation kinetics, resulting in $T_s = T_{sat}(p_H)$ or equivalently $\Bt=0$. Consequently, $\Lambda$ is independent of $\Kn$. However, this assumption starts to break down as $\Kn$ exceeds $O(10^{-1})$. With a further increase in $\Kn$, $J$ at $T_{sat}$ is not sufficient to dissipate the heat diffused into the interface, causing $T_s$ to rise above $T_{sat}$ (i.e.\, a positive $\Bt$, as shown in fig.\ref{fig2b}). This increase in $T_s$ leads to a reduction in $q_w$ due to the reduced temperature gradient across the thin liquid film which eventually leads to a reduction in $J$ and $\Lambda$ (see fig.\ref{fig2a}). Counting for the temperature jump across the Knudsen layer (i.e.\, red lines, $T_s > T_k \approx T_H = T_{sat}$) further contributes to an increase in $T_s$ or equivalently $\Bt$ and a subsequent reduction in $\Lambda$. The assumption of negligible $q_v$ (red lines, the one-sided model) does not significantly impact $\Bt$ (as depicted in fig.\ref{fig2b}) but does lead to an overestimation of $\Lambda$ (compared to the black line which is the most general full solution) due to the ignored heat loss as sensible heat into the vapour domain. Please note that all results are generated in cases where the top boundary is at the thermodynamic equilibrium (i.e.\, $T_H=T_{sat}(p_H)$), the impact of ignoring $q_v$ could be more significant in cases when $T_H > T_{sat}(p_H)$, i.e.\, superheated vapour. Fig.\ref{fig2} also shows that the linearised solutions provide good approximations to the full solution, especially when $\Kn \leq O(10^0)$, owning to the proximity of $\Bt$ to $0$. 
\begin{figure}[h]
\centering
\begin{subfigure}{.32\textwidth}
\centering
\includegraphics[width = \textwidth]{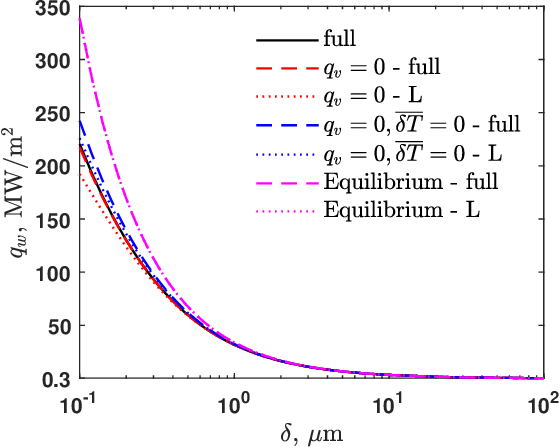}
\caption{ }
\label{fig3a}
\end{subfigure}
\hfill
\begin{subfigure}{.32\textwidth}
\centering
\includegraphics[width = \textwidth]{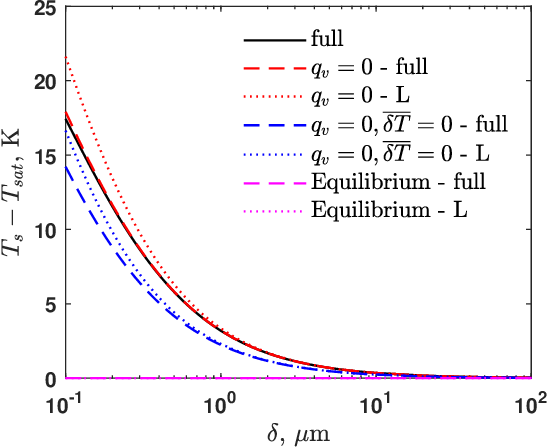}
\caption{ }
\label{fig3b}
\end{subfigure}
\hfill
\begin{subfigure}{.32\textwidth}
\centering
\includegraphics[width = 1.1\textwidth]{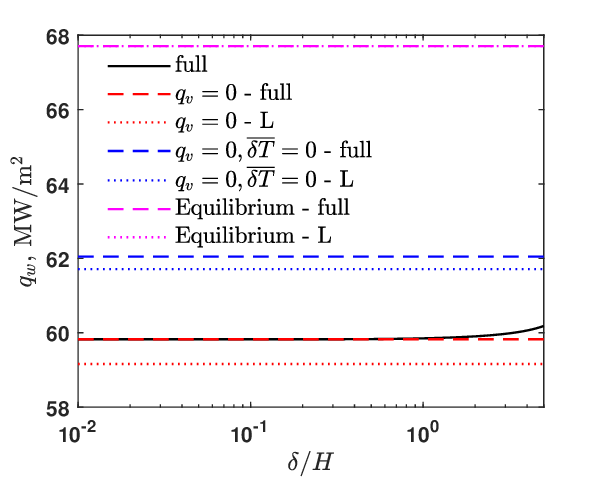}
\caption{ }
\label{fig3c}
\end{subfigure}
\caption{Impact of $\delta$ on: (a) $q_w$ and (b) $T_s-T_{sat}$, and (c) impact of changing $H$ on $q_w$, under different assumptions, including $q_v = 0$, $\bat = 0$ or $T_s=T_{sat}$. All results shown on (a) and (b) are obtained when $\badel=1$. All results shown on (c) are for $\delta=0.5\,\upmu$m. In all cases, $T_H = T_{sat}(p_H)=372.76 \mathrm{K}, p_H=1$ bar. `full' - full solutions; `L' - linearised solutions. }
\label{fig3}
\end{figure}

The thermal resistance ratio $R_{ig}/R_l$ at different $\Kn$ is shown in fig.\ref{fig2c}. As expected, the assumption of thermodynamic equilibrium means there is no thermal resistance across the interface. However, once the finite-rate evaporation kinetics and related temperature jump are counted for,  $R_{ig}/R_l$ exhibits a consistent increase with higher $\Kn$ values (e.g. reduced $\delta$). Once $\Kn \geq O(1)$, $R_{ig}$ can be on the same orders of magnitude to $R_l$. One would expect that $R_{ig}$ can surpass $R_{l}$ as $\Kn$ further increases beyond the range plotted in fig.\ref{fig2c}, making the interfacial thermal resistance the rate-limiting factors in dissipating heat. This could well be the case based on the model proposed here. However, the results will not be accurate quantitatively since the Casimir force, which can be significant as $\Kn$ goes beyond 10, has not been counted for. We would expect the qualitative trend still holds considering the Casimir force serves to further suppress the evaporation rate.

In practical applications, the concern lies in the achievable $q_w$ at different $\delta$ (fig.\ref{fig3a}) and $H$ (fig.\ref{fig3c}). As expected, fig.\ref{fig3a} shows that a reduction in $\delta$ increase the temperature gradient across the thin film so $q_w$ increases. In fig.\ref{fig3a}, we can also clearly see that ignoring the evaporation kinetics for a relatively thick film (e.g. $\delta \geq 1\,\upmu$m) does not introduce any noticeable error in $q_w$, but can lead to a significant overestimation in $q_w$ for a a thinner film. This overestimation occurs due to an underestimated $T_s$ (see fig.\ref{fig3b}) which subsequently overestimates the temperature gradient across the thin film. fig. \ref{fig3a} also shows that the linearised solutions provide good approximations to the full solutions when predicting $q_w$. The impact of changing $H$ on $q_w$ is shown in fig.\ref{fig3c}. For models disregarding $q_v$, $H$ is irrelevant to the problem; hence the straight lines. However, reducing $H$ can have a noticeable impact on $q_w$ when $\badel \geq 1$ once both $q_v$ and evaporation kinetics are counted for.

\subsection{Evaporation of water film in gaseous water-nitrogen mixture} \label{sec:Results-inert}
\begin{figure}[t]
\centering
\begin{subfigure}{.49\textwidth}
\includegraphics[width = .7\textwidth, right]{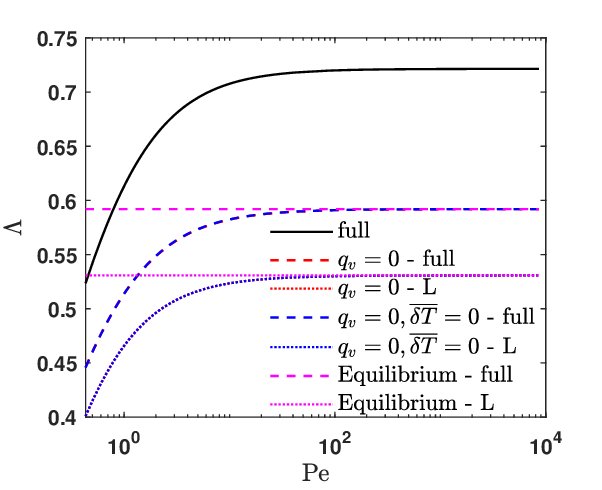}
\caption{ }
\label{fig4a}
\end{subfigure}
\hfill
\begin{subfigure}{.49\textwidth}
\includegraphics[width = .66\textwidth, left]{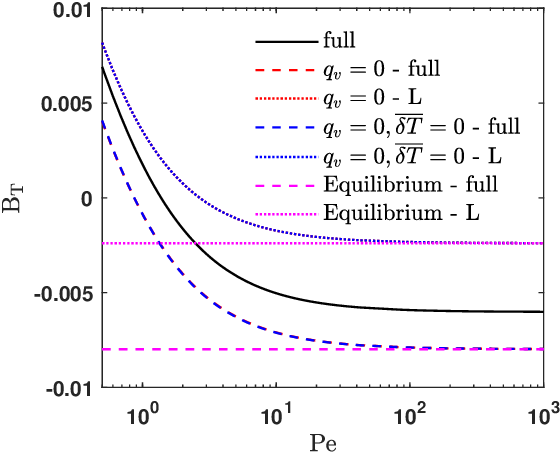}
\caption{ }
\label{fig4b}
\end{subfigure}
\hfill
\begin{subfigure}{.49\textwidth}
\includegraphics[width = .65\textwidth, right]{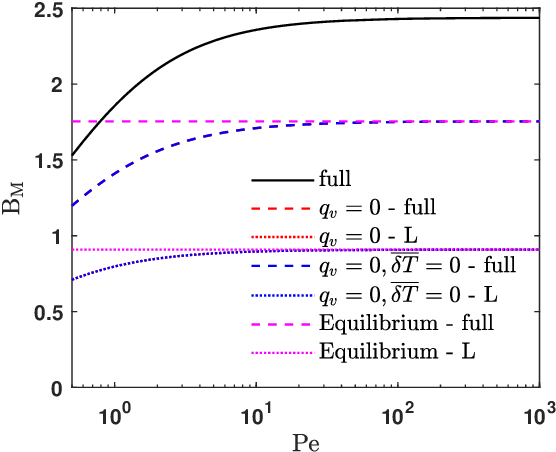}
\caption{ }
\label{fig4c}
\end{subfigure}
\hfill
\begin{subfigure}{.49\textwidth}
\includegraphics[width = .7\textwidth, left]{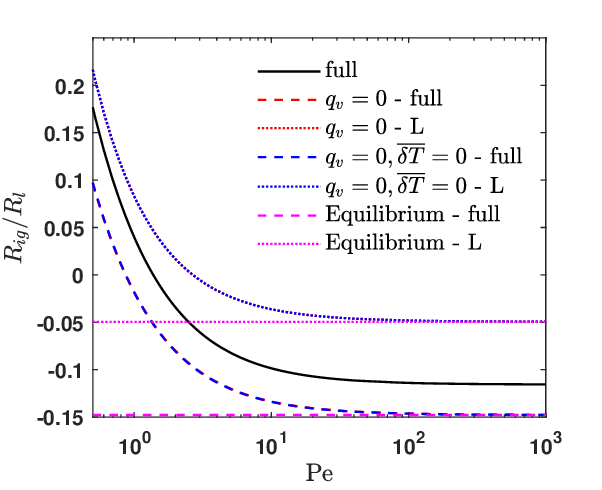}
\caption{ }
\label{fig4d}
\end{subfigure}
\caption{Impact of $\Pe$ on (a) $\Lambda$, (b) $\Bt$, (c) $\Bm$ and (d) $R_{ig}/R_l$, under different assumptions, including $q_v = 0$, $\bat = 0$ or $T_s=T_{sat}$. In all cases, $\badel=1, T_H = T_{sat}(p_H)=372.76 \mathrm{K}, p_H = 1$ bar, and $\Phi_H=1$ (i.e.\, pure nitrogen at the top boundary). `full' - full solutions; `L' - linearised solutions. Please note that the blue and red lines are mostly overlapped.}
\label{fig4}
\end{figure}
\begin{figure}[h]
\centering
\includegraphics[width = 0.33\textwidth]{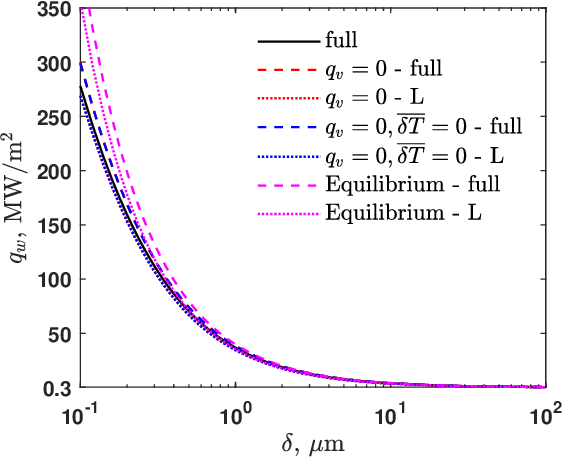}
\caption{Impact of $\delta$ on $q_w$ under different assumptions, including $q_v = 0$, $\bat = 0$ or $T_s=T_{sat}$. In all cases, $\badel=1, T_H = T_{sat}(p_H)=372.76 \mathrm{K}, p_H = 1$ bar, and $\Phi_H=1$ (i.e.\, pure nitrogen at the top boundary). `full' - full solutions; `L' - linearised solutions. Please also note that the blue and red lines are mostly overlapped.}
\label{fig5}
\end{figure}

The impact of different assumptions on the predicted $\Lambda, \Bt, \Bm$ and $R_{ig}/R_l$ at varying $\Pe$ is presented on fig.\ref{fig4a} - \ref{fig4d}. These plots reveal that the impact of finite-rate evaporation kinetics only becomes apparent when $\Pe$ drops below $O(10^2)$. The commonly assumed diffusion-dominated regime corresponds to the cases where $\Pe \rightarrow \infty$ but it remains a good approximation when $\Pe \geq O(10^2)$. Similar to evaporation in water vapour, counting for the evaporation kinetics results in a reduction in $\Lambda$ and an increase in both $\Bt$ and $R_{ig}/R_l$ for the same reason: $T_s$ needs to increase to dissipate the heat diffused into the interface. However, several different features, compared to the cases of evaporation in water vapour, can be observed from fig.\ref{fig4}. Firstly, ignoring $q_v$ can introduce more substantial errors, and the solutions with different assumptions do not converge for thick films (i.e.\, large $\Pe$). This is because $T_s$, which now strongly depends on the partial pressure of water at the interface (i.e.\, $p_s^A$), can be significantly lower than $T_H$ (set equal to $T_{sat}(p_H)$) when the mole fraction of water at the interface $x_s^A < 1$ (as evidenced by the finite $\Bm$ shown in fig.\ref{fig4c}: $\Bm \rightarrow \infty \,\mathrm{as}\, x_s^A \rightarrow 1$) so $p_s^A (= x_s^A p_H)< p_H$. This leads to a much stronger temperature gradient and $q_v$ in the gas phase. Secondly, fig.\ref{fig4a} shows that the assumption of negligible $q_v$ underestimates $\Lambda$ (or equivalently $J$ at constant $\delta$ and gas properties) when there are inert gases, contrasting cases of evaporation in pure vapours. This is again due to the fact that $T_s$ can be lower than $T_H$ (i.e.\, a negative $\Bt$ as shown in fig.\ref{fig4b}) when there are inert gases. Such a negative temperature gradient leads to additional heating to the interface to evaporate more liquid so ignoring this heat ($q_v$) will underestimate $J$ and $\Lambda$. Lastly, the linearised solutions do not offer good approximations to the full solutions, mainly because $\Bm$ can significantly exceeds 0. However, the small value of $\Bt$ raises the hope that the linearised solution for $T_s$ and consequentially $q_w$ might offer a decant approximation to the full solution. This is indeed the case, as shown in fig.\ref{fig5}.

\begin{figure}[t]
\centering
\begin{subfigure}{.32\textwidth}
\centering
\includegraphics[width = \textwidth]{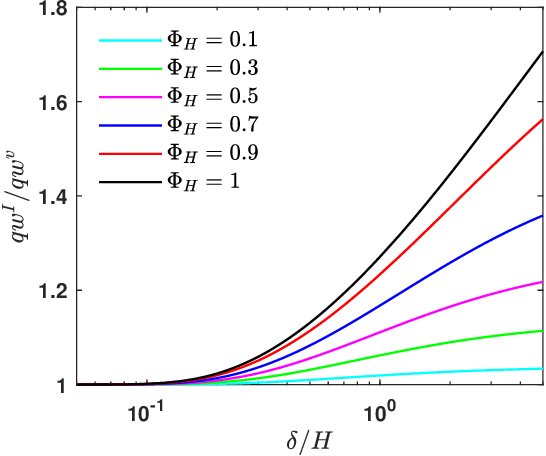}
\caption{$\delta = 1$ mm}
\label{fig6a}
\end{subfigure}
\hfill
\begin{subfigure}{.32\textwidth}
\centering
\includegraphics[width = \textwidth]{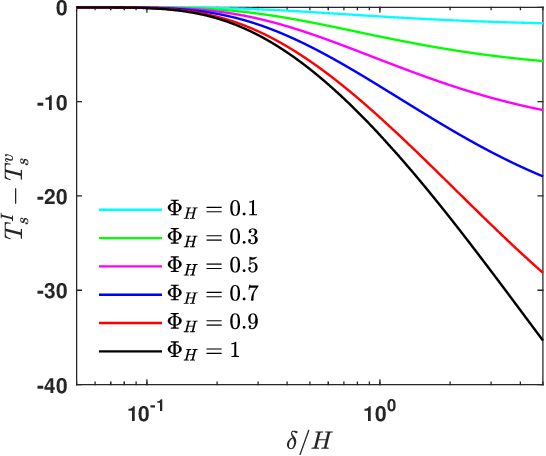}
\caption{$\delta = 1$ mm}
\label{fig6b}
\end{subfigure}
\hfill
\begin{subfigure}{.32\textwidth}
\centering
\includegraphics[width = \textwidth]{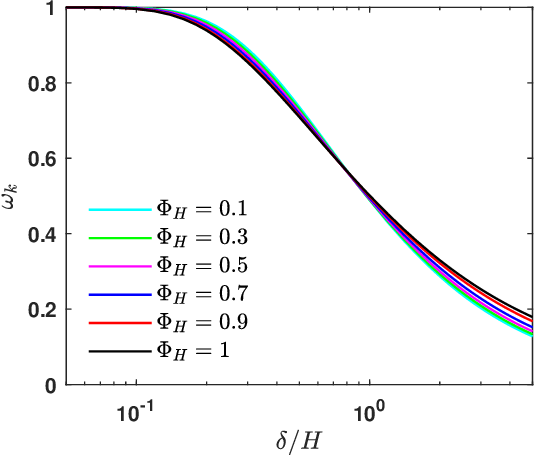}
\caption{$\delta = 1$ mm}
\label{fig6c}
\end{subfigure}
\begin{subfigure}{.32\textwidth}
\centering
\includegraphics[width = \textwidth]{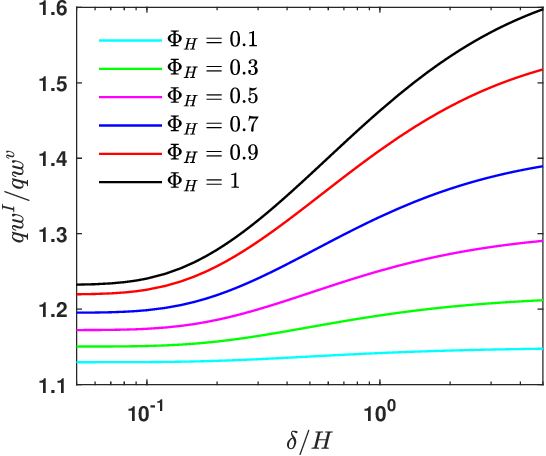}
\caption{$\delta = 0.1\,\upmu$m}
\label{fig6d}
\end{subfigure}
\hfill
\begin{subfigure}{.32\textwidth}
\centering
\includegraphics[width = \textwidth]{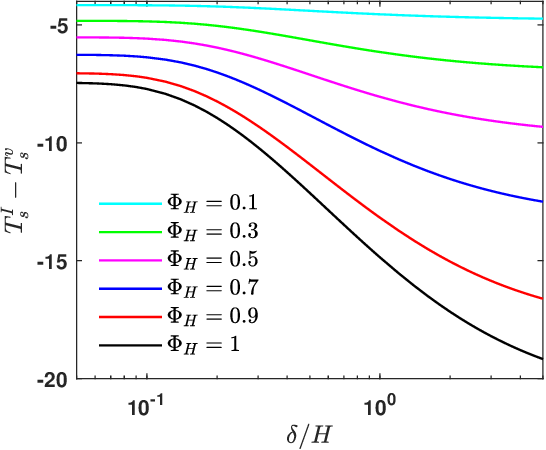}
\caption{$\delta = 0.1\,\upmu$m}
\label{fig6e}
\end{subfigure}
\hfill
\begin{subfigure}{.32\textwidth}
\centering
\includegraphics[width = \textwidth]{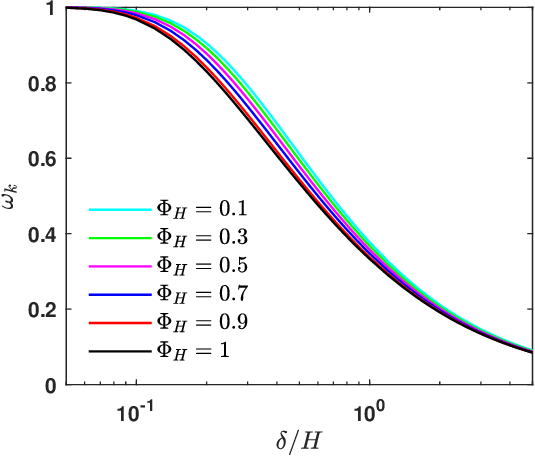}
\caption{$\delta = 0.1\,\upmu$m}
\label{fig6f}
\end{subfigure}
\caption{Impact of $\badel$ and $\Phi_H$ on: (a) \& (d) $q_w^I/q_w^v$, (b) \& (e) $T_s^I-T_s^v$, and (c) \& (f) $\omega_k$. Superscripts $I$ and $v$ refers to cases of evaporation in inert gases and in pure vapours respectively. All results are from full numerical solutions to the general problem (i.e.\, eqs.\ref{eq:BC2_g_D}, \ref{eq:BC3_g_D}, \ref{eq:Dg_s} and \ref{eq:Eg_s}). Results shown on (a)-(c) are for $\delta=1$ mm so the problem is diffusion dominated while (d)-(e) are for $\delta = 0.1\,\upmu$m where the kinetic effect is significant. In all cases, $p_H = 1$ bar and $T_H=T_{sat}(p_H)=372.76$ K. Please also note that $q_w^v$ are calculated based on evaporation in pure vapours with $p_H=1$ bar and $T_H = 372.76$ K.}
\label{fig6}
\end{figure}

When evaporating in inert gases, another parameter which can significantly affect the evaporation rate is the mass fraction distributions of inert gases in the gas domain. While the evaporation of liquid at the interface is strongly dependant on the finite-rate evaporation kinetics in a pure vapour environment, in inert gases, the rate-limiting factor will be the mass diffusion rate adjacent to the interface (see eq.\ref{eq:BC2_g}). This has been evidenced both in experiments \citep{Lu2017} and modelling results shown in fig.\ref{fig4}, especially for thick film (i.e.\, $\Pe \geq O(10^2)$). The mass diffusion rate depends on $H$ and $\Phi_H$, and the influence of these two parameters on the evaporation process is illustrated in fig.\ref{fig6}. 

For thick film (i.e.\, diffusion-dominated evaporation), it is evident from figs.\ref{fig6a}-\ref{fig6c} that evaporation in a pure vapour environment can be considered as the limiting case when either $\Phi_H \rightarrow 0$ (fig.\ref{fig6a}) or $\omega_k \rightarrow 1$ (fig.\ref{fig6c}). In both cases, the interface temperature in inert gases $T_s^I \rightarrow T_{sat}(p_H)$ which is the same as the interface temperature in pure vapours $T_s^v$ (as shown in fig.\ref{fig6b}) such that $q_w$ will be the same (i.e.\, $q_w^I/q_w^v = 1$ where the superscripts $I$ and $v$ refers to the cases of inert gases and pure vapours respectively, see fig.\ref{fig6a}). When inert gases present at the interface (i.e.\, $\omega_k < 1$), the partial pressure of the water vapour $p_s^A$ becomes less than $p_H$ so that the corresponding $T_s \approx T_{sat}(p_s^A) < T_{sat}(p_H)$ (please note that the kinetic effect is negligible for thick film and the correction to the $T_{sat}$ due to inert gases is also negligibly small at such a low total pressure $p_H$, i.e.\, $\bavs \approx 0$ \citep{Zhao2023}). This leads to a larger temperature gradient across the liquid film hence a larger $q_w$. To harness the higher cooling capacity in inert gases compared to a pure vapour environment due to reduced $p_s^A$ at the interface, however, require the gas domain thickness $H$ to be comparable or smaller than the film thickness (i.e.\, $\badel \leq O(10^{-1})$) while maintaining the $\Phi_H$ at the top boundary.

For thin films (i.e.\, evaporation kinetics play significant roles), fig.\ref{fig6e} shows that $T_s^I$ will be consistently lower than $T_s^v$, resulting in consistently higher $q_w^I$ than $q_w^v$ (fig.\ref{fig6d}), even when $\omega_k \rightarrow 1$ (fig.\ref{fig6f}). This is due to the additional role the inert gases play in thermalizing the vapour molecules emerging from the liquid. Collisions between the inert gases and the vapour molecules lead to additional energy exchange, causing a significantly lower temperature jump across the Knudsen layer in inert gases compared to that in pure vapours. Such a crucial role of inert gases is captured by the temperature jump model shown in eq.\ref{eq:BC3_g}. However, this model is empirical in nature and should be derived more rigorously by counting for both the temperature jump caused by heat/mass exchange across the interface, as in eq.\ref{eq:R13}, and the thermalisation process due to inert gases. One would expect that eq.\ref{eq:R13_1D} will be the limiting case of a more general temperature jump model where the collision-induced thermalisation process is negligibly small. Therefore, the non-zero $T_s^I - T_s^v$ (which leads to non-unity $q_w^I/q_w^v$) as $\Phi_H \rightarrow 0$ in fig.\ref{fig6a} and \ref{fig6b} highlights the necessity for significant correction of existing temperature jump model in inert gases for achieving consistency.
\begin{figure}[t]
\centering
\begin{subfigure}{.32\textwidth}
\centering
\includegraphics[width = \textwidth]{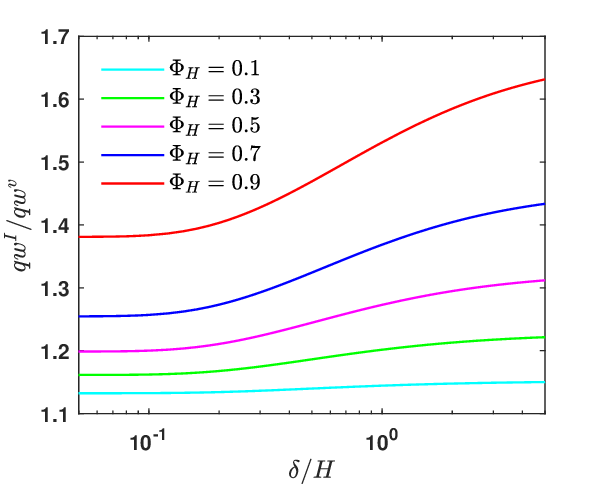}
\caption{$\delta = 0.1\,\upmu$m}
\label{fig7a}
\end{subfigure}
\hfill
\begin{subfigure}{.32\textwidth}
\centering
\includegraphics[width = \textwidth]{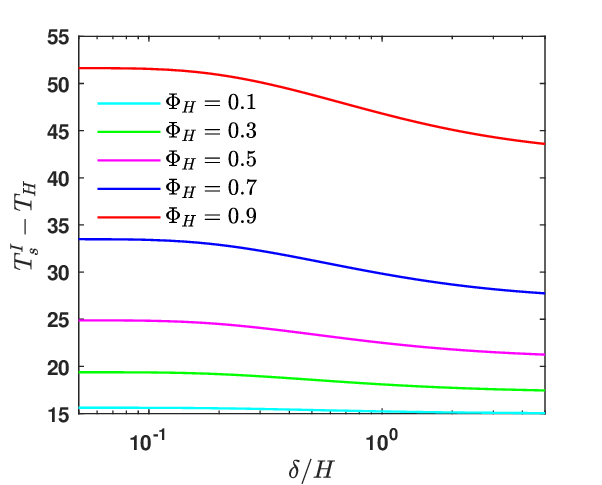}
\caption{$\delta = 0.1\,\upmu$m}
\label{fig7b}
\end{subfigure}
\hfill
\begin{subfigure}{.32\textwidth}
\centering
\includegraphics[width = \textwidth]{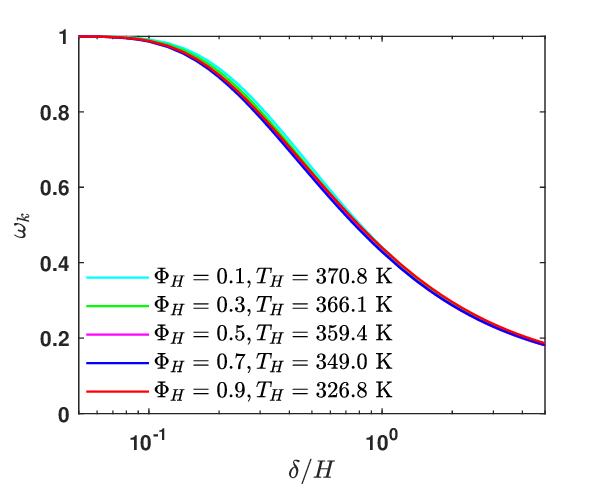}
\caption{$\delta = 0.1\,\upmu$m}
\label{fig7c}
\end{subfigure}
\caption{Impact of $\badel$ and $\Phi_H$ on: (a) $q_w^I/q_w^v$, (b) $T_s^I-T_H$, and (c) $\omega_k$. All results shown on (a) and (b) are from full numerical solutions to the general problem (i.e.\, eq.\ref{eq:BC2_g_D}, \ref{eq:BC3_g_D}, \ref{eq:Dg_s} and \ref{eq:Eg_s}). In all cases, $p_H = 1$ bar and $T_H = T_{sat}(p_H^A)$. Please also note that $q_w^v$ are calculated based on evaporation in pure vapours with $p_H=1$ bar and $T_H = 372.76$ K.}
\label{fig7}
\end{figure}

In practice, controlling $\Phi_H$ can be achieved by adjusting $T_H$ and allowing the top boundary to be saturated with the water vapour (i.e.\, $p_H^A = p_{sat}(T_H)$ or equivalently the relative humidity reaches 1). Any additional water reaching the top boundary will then be condensed. It is clear from fig.\ref{fig7a} that this strategy leads to a significant reduction in both $T_H$ and $T_s$ and reinstating a positive $\Bt$ (equivalently $T_s^I > T_H$ as shown in fig.\ref{fig7b}). This alternation causes additional heat to be taken away from the interface to the top boundary. Consequently, there is a notable increase in $q_w^I$ compared to the strategy of maintaining $T_H$ at $T_{sat}(p_H)$ (see results on fig.\ref{fig6}). While $q_w^I$ is increased, the reduction in $T_s^I$ actually diminish $J$, thereby reducing $\omega_k$ (as observed in the comparison between fig.\ref{fig7c} and fig.\ref{fig6e}). 

\begin{figure}[t]
\centering
\begin{subfigure}{.32\textwidth}
\centering
\includegraphics[width = \textwidth]{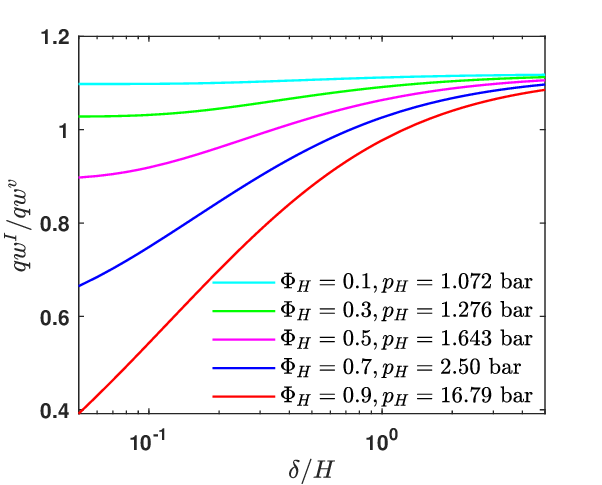}
\caption{$\delta = 0.1\,\upmu$m}
\label{fig8a}
\end{subfigure}
\hfill
\begin{subfigure}{.32\textwidth}
\centering
\includegraphics[width = \textwidth]{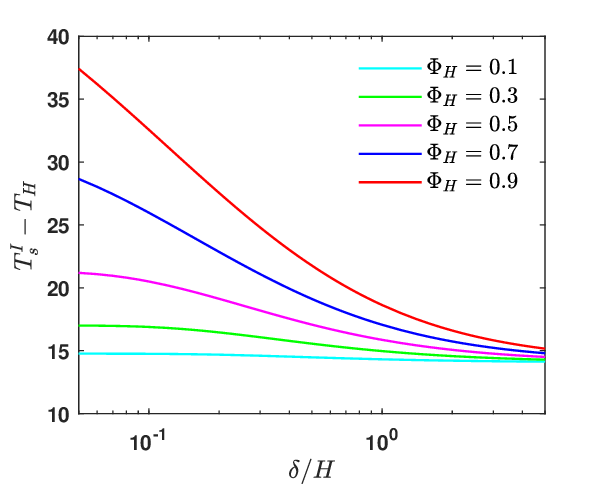}
\caption{$\delta = 0.1\,\upmu$m}
\label{fig8b}
\end{subfigure}
\hfill
\begin{subfigure}{.32\textwidth}
\centering
\includegraphics[width = \textwidth]{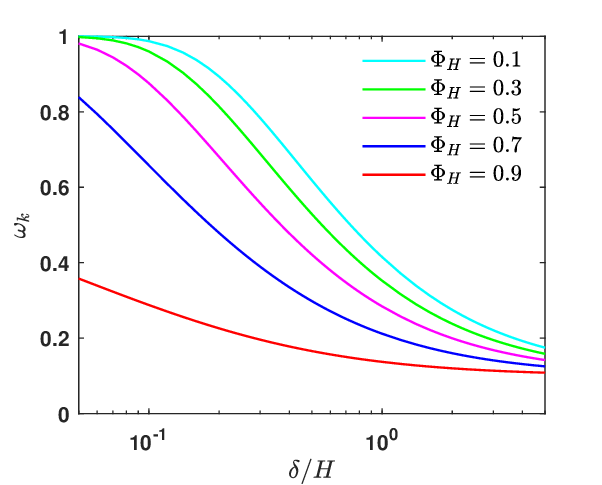}
\caption{$\delta = 0.1\,\upmu$m}
\label{fig8c}
\end{subfigure}
\caption{Impact of $\badel$ and $\Phi_H$ on: (a) $q_w^I/q_w^v$, (b) $T_s^I-T_H$, and (c) $\omega_k$. All results shown on (a) and (b) are from full numerical solutions to the general problem (i.e.\, eq.\ref{eq:BC2_g_D}, \ref{eq:BC3_g_D}, \ref{eq:Dg_s} and \ref{eq:Eg_s}). In all cases, $T_H = 372.76$ K and $p_H^A = p_{sat}(T_H)$. Please also note that $q_w^v$ are calculated based on evaporation in pure vapours with $p_H=1$ bar and $T_H = 372.76$ K.}
\label{fig8}
\end{figure}

One can also ask the questions that what the impact is to keep $T_H = T_{sat}(p_H^A)$ as a constant while changing the $\Phi_H$ by increasing the amount of inert gases (i.e.\, increasing $p_H$). Fig.\ref{fig8a} shows that increasing $p_H$ in this case will substantially reduce the achievable $q_w$ at the same $T_w$. This understandably is due to the significant increase of the $T_s$ (see fig.\ref{fig8b}), resulting from an increase in $p_s^A$. Meanwhile the high saturation temperature corresponding to $p_s^A$ leads to a drastic reduction in $J$, consequently diminishing $\omega_k$ (see fig.\ref{fig8c}).
\section{Conclusion} \label{sec:Conclusion}
This paper presents a theoretical analysis addressing the general problem of evaporation of a flat thin liquid film in either pure vapours or vapour-inert-gas mixtures, incorporating the finite-rate evaporation kinetics which are typically overlooked in most existing literature. The main objective of this analysis is to find out whether evaporation kinetics can be the rate-limiting factors during thin-film evaporative heat transfer processes. This is achieved by establishing a thermal resistance network encompassing the whole thermal path: from the flat hot surface, through the liquid domain, across the interface region and finally through the gas domain. The thermal resistance in the liquid domain $R_l$ and that across the interface and the gas domain $R_{ig}$ are then quantified by analysing the heat and mass transport processes using macroscopic transport equations for continuum fluids which are coupled together with kinetic boundary conditions. When analysing the evaporation of thin film in pure vapours, the kinetic boundary conditions developed by Struchtrup et al. \citep{Struchtrup2017}, which counts for both the finite evaporation mass flux and temperature jump across the out-of-equilibrium Knudsen layer, is used. For problems involving evaporation in vapour-inert-gas mixtures, rigorous kinetic models which counts for the interaction between vapour and inert gases are not readily available. Hence, the generalised Hertz-Knudsen correlation with correction to the change of chemical potential of the vapour phase due to inert gases \citep{Zhao2023} together with an empirical temperature jump correlations based on fittings to molecular dynamic simulation results for evaporation of water drops in nitrogen \citep{Hoyst2013a} are used. Through non-dimensionlisation, the key dimensionless parameters to determine the importance of evaporation kinetics are identified to be $\Kn$ in pure vapour environment and $\Pe$ in inert gases. Full numerical solutions and analytical asymptotic solutions at the slow evaporation limit are then developed within these dimensionless frameworks. These solutions are subsequently applied to analyse the evaporation of a thin water film evaporating in either pure water vapours or a water-nitrogen gas mixtures.

The modelling results show that:
\begin{itemize}
\item The widely adopted assumptions of negligible temperature jump and a thermodynamically equilibrial interface are only strictly valid in the limit of $\Kn \rightarrow 0$ or $\Pe \rightarrow \infty$. However, these two assumptions can provide good approximations when $\Kn \leq O(10^{-1})$ or $\Pe \geq O(10^2)$, which corresponding to the evaporation of a water film thicker than a few $\upmu$m. In this regime, $R_{ig}$ will be negligibly small.
\item For sufficiently thin film (i.e.\, $\Kn \geq O(10^{-1})$ in a pure vapour environment or $\Pe < O(10^2)$ in inert gases), $R_{ig}$ attributable to finite-rate evaporation kinetics becomes increasingly more significant as the film gets thinner. $R_{ig}$ will be on the same orders of magnitude to $R_l$ when $\Kn \leq O(1)$ or $\Pe \geq O(1)$.
\item The widely adopted assumption of negligible $q_v$ provide good approximations only when $T_H \approx T_{sat}(p_s^A)$ and only for relative thick film (i.e.\, when the evaporation kinetics is infinitely fast). 
\item Simplified explicit analytical asymptotic solutions can offer very good approximations when \{$\Bt, \bat$\} are close to 0 in predicting $q_w$, a common scenario in practical applications. However, $\Bm$ can be significantly larger than 0 in practice such that the error in predicting $J$ can be significant.
\item For relative thick film (i.e.\, when the evaporation kinetics is infinitely fast), the evaporation of a liquid film in pure vapours can be considered as the limiting cases of the more general problems of evaporation in vapour-inert-gas mixtures when either $\Phi_H \rightarrow 0$ or $\omega_k \rightarrow 1$. Increasing $\Phi_H$ can effectively reduce $T_s$ and subsequently enhance $q_w$ when $\badel \leq 0.1$.
\item For thin film where the evaporation kinetics becomes less efficient, $q_w$ during evaporation in inert gases is always larger than that in pure vapours because the inert gases can serve to reduce the temperature jump and thus $T_s$.
\item The strategy of increasing $\Phi_H$ by reducing $T_H$ while maintaining $p_H$ in a 100\% humid top boundary can increase $q_w$ effectively by lowering $T_s$.
\item Conversely, the strategy of increasing $\Phi_H$ by increasing $p_H$ while maintaining $T_H$ in a 100\% humid top boundary will substantially reduces $q_w$ by elevating $T_s$.
\end{itemize}

Meanwhile, the modelling results also exposed a significant knowledge gap which needs to be addressed: the lack of a rigorous temperature jump model in vapour-inert-gas mixtures. Such a temperature jump model will need to incorporate both the energy exchange during collisions of inert gases and vapour molecules but also the change of enthalpy across the interface and out-of-equilibrium kinetic layer due to heat and mass transport processes.

\section*{Conflict of interest}
The authors declare that there is no conflict of interest.

\bibliographystyle{unsrt}
\bibliography{IJHMT-2024}

\end{document}